\documentclass{pasj00}
\draft

\begin{document}
\SetRunningHead{Author(s) in page-head}{Running Head}
\Received{2003/01/16}%{yyyy/mm/dd}
\Accepted{2003/04/16}%{yyyy/mm/dd}

\title{Detection of Molecular Clouds in the Interarm of the Flocculent Galaxy 
NGC 5055}

%%% begin:list of authors
%\author{Tomoka \textsc{Tosaki}%,
%  \thanks{Example: Present Address is xxxxxxxxxx}}
%\affil{Gunma Astronomical Observatory, Nakayama, Takayama, Agatsuma,
%Gunma, 377-0702, Japan} \\
%{\it 377-0702, Japan} \\
%\email{tomoka@astron.pref.gunma.jp}
 
%\author{Yasuhiro \textsc{Shioya}},
%\affil{Tohoku University, Aoba, Sendai, Miyagi, 980-8578, Japan}
%\email{shioya@astr.tohoku.ac.jp},
%\and
%\author{Nario {\sc Kuno}}
%\affil{Nobeyama Radio Observatory, Minamimaki, Minamisaku, Nagano, 
%384-1805, Japan}
%\email{kuno@nro.nao.ac.jp}

%%% end:list of authors

%%% Please use the following style in case that sorting by 
%%% affilation is impossible. 
%
 \author{%
   Tomoka \textsc{Tosaki},\altaffilmark{1}
   Yasuhiro \textsc{Shioya},\altaffilmark{2} \\
   Nario \textsc{Kuno},\altaffilmark{3}
   Kouichiro \textsc{Nakanishi},\altaffilmark{3}
   and 
   Takashi \textsc{Hasegawa}\altaffilmark{1} }
 \altaffiltext{1}{Gunma Astronomical Observatory, Nakayama, Takayama, 
                  Agatsuma, Gunma 377-0702}
 \email{tomoka@astron.pref.gunma.jp}
% \email{eeeee@xxx.xxx.xx.xx}
 \altaffiltext{2}{Tohoku University, Aoba, Sendai, Miyagi 980-8578}
 \altaffiltext{3}{Nobeyama Radio Observatory, Minamimaki, Minamisaku, Nagano 
                  384-1805}

%% `\KeyWords{}' always has to be placed before `\maketitle'.
\KeyWords{galaxies: ISM --- galaxies: spiral --- galaxies: individual (NGC 5055)} %Do NOT move this preamble from here!

\maketitle

\begin{abstract}

 We present high-resolution ($\sim \timeform{4''}$) $^{12}$CO ($J$~=~1--0) 
mapping observations with high-velocity resolution ($\sim 2.6~$km s$^{-1}$)
toward the disk of flocculent galaxy NGC 5055, 
using the Nobeyama Millimeter Array in order to study the physical properties 
of the molecular clouds in the arm and the interarm.
The obtained map shows clumpy structures.
Although these are mainly distributed along a 
spiral arm seen in near-infrared observations, 
some clouds are located far from the arm, namely in the interarm.
These clouds in both the arm and the interarm have a typical size and mass 
 of a few 100 pc and a few 10$^6M_\odot$, respectively. 
These correspond to the largest Giant Molecular Cloud (GMC) 
in our Galaxy, and are slightly
smaller than Giant Molecular Associations (GMAs) in the grand design spiral M 51.
Their CO flux-based masses show good agreement with their virial masses.
A size--velocity dispersion relation is also plotted on an extension of 
the relation for the Galactic GMCs.
These facts suggest that the properties of these clouds are similar to
that of the Galactic GMCs. 
%We found no significant difference in the GMC properties, such as the mass
%and size, between the arm and the interarm unlike M 51. 
%The lifetime of the interarm GMCs were estimated to be 
%similar to the timescale after leaving arm.
%This suggests that local fluctuation of gas density can form GMCs
%in the interarm as well as the arm. 
We also found no clear systematic offset 
between the molecular gas and H\,{\footnotesize II} regions unlike M 51. 
This fact and no existense of GMAs suggest the view that, in NGC 5055, cloud formation 
and following star formation in both the arm and the interarm are due to enhancement 
of gas by local fluctuation.
%A larger scale structure, like GMA, which is considered to 
%be formed by a large-scale gravitational
%instability and/or cloud collision, could not seen in NGC 5055.
%We think that this is because the density enhancement in the arm
%of NGC 5055 is not enough, since the density wave is weak.
On the other hand, in grand design spiral galaxies, such as M 51, 
%although we could see GMCs both in the arm and the interarm,  
GMA formations may occur only in the arm due to a strong density wave 
also enhanced star formation in GMA formation may also occur.    
These may control the optical morphology of spiral arms in spiral galaxies.

\end{abstract}

\section{Introduction}

Spiral structures are a striking feature of galaxies, and
are considered to have a correlation with dense gas formation 
and/or star formation
because spiral structures trace the density wave that is expected to 
accumulate or compress molecular gas. 
We can find  various morphological types of 
spiral arms in many galaxies.
\citet{key-Elme82} and \citet{key-Elme87} introduced the idea of an arm class 
as a classification system 
of spiral arms based on the arm continuity, length, and symmetry, and
this may be related to the existence or strength of a density wave.  
The arm classes have a range from 1 to 12, and are classified into 
3 main groups, i.e., flocculent, multiple, and grand design. 
Flocculent spiral galaxies have short and patchy spiral arms, whereas
grand design spiral galaxies show long and continuous spiral arms.
Multiple spiral galaxies are located at intermediate type between them.

For the grand design spiral galaxy of M 51, 
an offset among $^{12}$CO, $^{13}$CO,
and H$\alpha$ emissions in spiral arms was found, namely, 
the $^{13}$CO and H$\alpha$ emissions were located at the downstream side of
$^{12}$CO (\cite{key-Tosa02}; \cite{key-Vog88}).
These offsets suggest that there is a time delay between an accumulation 
of gas caused by the density wave and dense gas formation and a following 
star formation.

On the other hand, recent near-infrared and CO observations showed that 
there are long, continuous spiral arms in the flocculent galaxy NGC 5055, 
though they are weaker than those of grand design spiral arms, such as M 51
(\cite{key-Kuno97a}; \cite{key-Thorn97}).
They suggest the possibility that density waves of old stars exist in spiral
galaxies universally, even if there is a difference in the strength.
This possibility indicates the view that a density wave plays a role for dense gas
and star formation in spiral galaxies.
However, only a few attempts have so far been made to perform high-resolution 
observations of molecular gas in the galactic disk, particularly, 
for flocculent galaxies. 
In order to investigate the role of density waves, we need information 
about the molecular 
cloud properties and star formation in spiral galaxies with different 
arm classes.

To study the molecular cloud properties in and between spiral arms 
and to consider the effect of spiral arms 
on the properties of molecular clouds, 
we present the results of observations with $^{12}$CO($J$~=~1--0) emission
toward the inner disk of NGC 5055.
NGC 5055 is a nearby flocculent spiral galaxy, which
has been investigated by several observations with various wavelength:
e.g., optical, infrared, radio, and so on.
This galaxy shows two weak spiral arms in near-infrared and CO
observations, 
and a kinematic indication of the density wave has been found in the galaxy.
Previous studies indicated the existence of large molecular clouds
in the central region of NGC 5055,
such as GMAs seen in nearby grand-design spiral M 51 (\cite{key-Thorn97}).
They found the GMAs on the spiral arm, but not between arms, e.g., interarm.

However, molecular clouds in the disk region of galaxies are supposed 
to have a smaller
velocity dispersion if they are similar to Galactic GMCs.
Therefore, if observations with a higher velocity resolution are carried out,
it is expected that we can detect interarm clouds.  
In consideration of this possibility, we performed observations 
toward the disk region of NGC 5055 with a velocity resolution of 
2.6 km s$^{-1}$, which is very high in extragalactic observations. 
We present the results as following sections.

The parameters of NGC 5055 are summarized in table \ref{tab:1}.

%\begin{verbatim}
\begin{table}
  \caption{Adopted parameters of NGC 5055.}\label{tab:1}
  \begin{center}
    \begin{tabular}{ll}
 \hline\hline
  Parameter & Value \\
\hline
  Center position (2000.0)$^*$ & 13$^{\rm h}$15$^{\rm m}$49.25$^{\rm s}$ \\
                           & 42$^\circ$01$^\prime$49.3$^{\prime\prime}$\\
  Morphological type$^\dag$ & SA(rs)bc \\
  Systemic Velocity$^\ddag$  & 504 km s$^{-1}$ \\
  Distance$^\ddag$ & 7.7 Mpc \\
  Position angle$^\S$ & 103$\pm$ 6$^\circ$ \\
  Inclination$^\S$ & 58$\pm$ 2$^\circ$  \\
\hline
{\footnotesize *: \cite{key-MA96}} \\
{\footnotesize \dag: \cite{key-dV91}} \\
{\footnotesize \ddag: \cite{key-WE86}} \\
{\footnotesize \S: \cite{key-GG91}} 
    \end{tabular}
  \end{center}
\end{table}

%\newpage

\section{Observations}

\begin{figure}
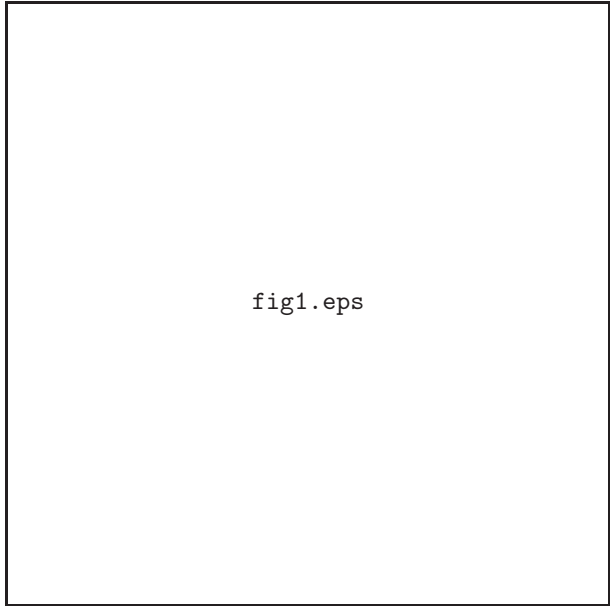

  \begin{center}
    \FigureFile(80m,80m){fig1.eps}
    %%% \FigureFile(width,height){filename}
  \end{center}
  \caption{Observed field on the optical image ($V$-band). The optical
 image was obtained by Subaru telescope equipped with a Suprime-Cam
(\cite{key-Komiya00}).
   }\label{fig:1}
\end{figure}

We carried out aperture synthesis observations of an arm of NGC 5055
in the $^{12}$CO($J$~=~1--0) emission line (rest frequency = 115.271204 GHz) 
using
the Nobeyama Millimeter Array (NMA) during 1999 November -- 2000 May.
The NMA consists of six 10 m antennas, equipped with SIS receivers.
%(\cite{key-Suna95}).
We show the observed field superposed on an optical image of NGC 5055 
in figure \ref{fig:1}.
As a back end, we used a 1024 channel FX spectrocorrelator
with a total bandwidth of 32 MHz, 
corresponding to 83.2 km s$^{-1}$ at the $^{12}$CO frequency.
The bandpass calibration was done with 3C 279, and 1216+487 was 
observed every 25 minutes as a phase calibrator. 
The flux scale of 1216+487 was determined by comparisons with 
planets of known brightness temperature. 
The uncertainty in the absolute flux scale is estimated to be $\sim$20\%.

The data were reduced using the NRO software package UVPROCII
(\cite{key-tsutsu97}), 
and the final maps were made and CLEANed with the NRAO software 
package AIPS. 
The size of the synthesized beam is \timeform{4.9''} $\times$ 
\timeform{3.6''}, corresponding to about 181 pc $\times$ 133 pc 
in linear scale at the distance of NGC 5055.
Velocity-channel maps were made with a velocity width of 2.6 km s$^{-1}$.
This is a very high-velocity resolution among observations toward 
external galaxies, and is effective to detect molecular clouds in 
disk regions, because molecular clouds in a disk are expected to
have a smaller velocity width compared with those in the central regions of
galaxies.

We measured the flux at a position near to the center of F.O.V, 
($\delta$R.A., $\delta$Decl)=($-$\timeform{30''}, $-$\timeform{28''}), 
which is the same position as the observed points by 45 m telescope,
to estimate the missing flux of our interferometric observations.
We deconvolved to the same resolution, \timeform{17''}, 
as the 45 m data (\cite{key-Kuno97a}).
The measured flux of NMA, 67 Jy km s$^{-1}$, is consistent with the 
45 m data, 65 Jy km s$^{-1}$, 
considering the errors of the flux scale, $\sim$ 20\%.
The line profile was also very similar to that by the 45 m data.
This indicates that the most of the single-dish flux has been recoverd by
our interferometric observations.

\section{Results}

\subsection{Global Distribution and Velocity Field of $^{12}$CO Emission}

\begin{figure}
  \begin{center}
    \FigureFile(80m,80m){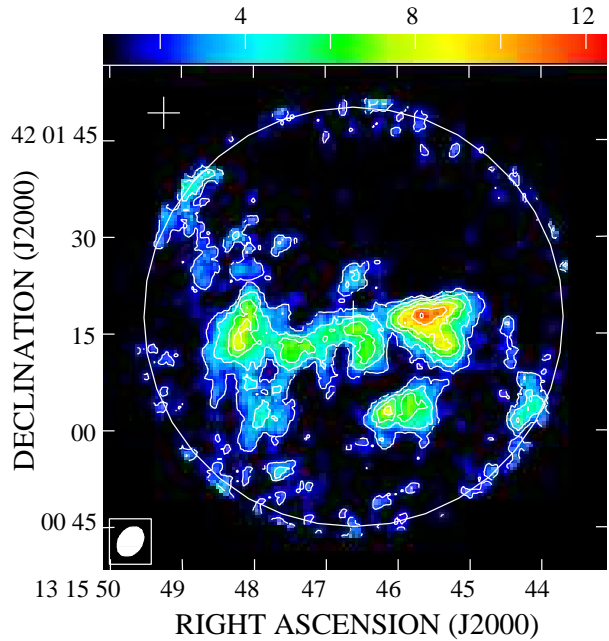}
    %%% \FigureFile(width,height){filename}
  \end{center}
  \caption{Total integrated intensity map of the $^{12}$CO emission 
   with NMA. The contour interval and lowest contour are 
   1 Jy km s$^{-1}$beam$^{-1}$, corresponding to noise level.
   The crosses at the top-left and center indicate the galactic center
  and center of F.O.V.
  The F.O.V is indicated as a circle and synthesized beam is plotted at 
  the bottom-left.
   }\label{fig:2}
\end{figure}

Figure \ref{fig:2} shows the total integrated intensity map of $^{12}$CO
emission over a velocity range of 82.3 km s$^{-1}$.
We found that the emission is distributed mainly along the spiral arm seen 
in near-infrared observations (\cite{key-Thorn97}).
The distribution of CO emission  consists of 
several clumps with a typical size of a few 100 pc.
These clumpy structures seem to be similar to that in a grand
design spiral galaxy M 51, which shows clear molecular spiral arms composed 
of GMAs (\cite{key-Rand90}; \cite{key-Tosa94a}a).
However, their sizes are smaller than the GMAs seen in M 51
whose size and mass are 1 kpc and $10^{7--8} M_\odot$, respectively.
We also must note that there are clumps outside of the spiral arm, 
namely, the interarm.
These clumps have the same sizes and intensities as those in the arm. 
This is unlike M 51. 
Although M 51 has interarm GMAs as well as on-arm GMAs, the emissions 
of the interarm GMAs are 
weaker and also are relatively smaller than the on-arm GMAs 
(\cite{key-Rand90}).
Most of clouds in the interarm of M51 are smaller than GMAs and 
have masses similar to those of GMCs in our Galaxy (\cite{key-Tosa94b}b).

\begin{figure}
  \begin{center}
    \FigureFile(80mm,80mm){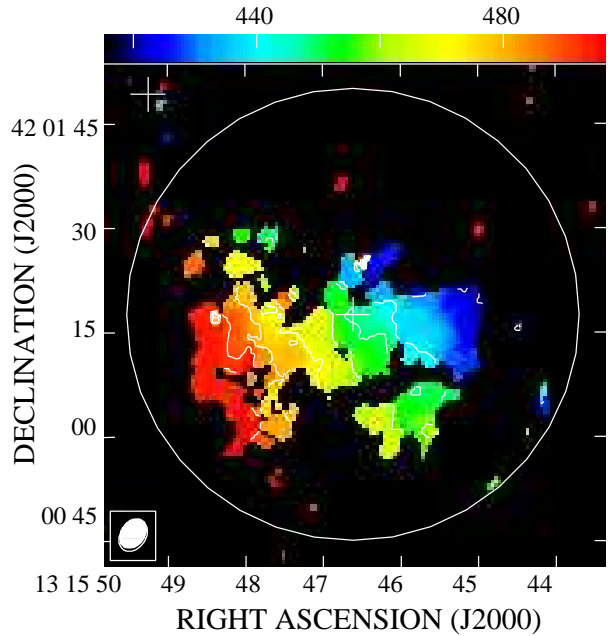} 
\end{center}
  \caption{Velocity field. Contour interval is 10 km s$^{-1}$.
The crosses and the circle indicate the same as in figure \ref{fig:2}.}
\label{fig:3}
\end{figure}

In figure \ref{fig:3}, we present the velocity field of the observed region 
which traces peaks of CO emission. 
This figure shows a systematic velocity field derived by
galactic rotation in both of the arm and the interarm.
If there exists a streaming motion caused by a density wave,
we can expect a distortion in the velocity field, such as an S-shape seen in 
the grand design spiral M 51 with 
a streaming motion of 50 -- 60 km s$^{-1}$ (\cite{key-Kuno97}; 
\cite{key-Aalt99}). 
However, we can not see such a distortion in NGC 5055.
In other words, we can not find any clear indicator of streaming motion 
caused by density waves.
This indicates that the effect of a density wave on
the kinematics in the arm of NGC 5055 is significantly weaker than that
in the grand design spiral M 51, even if one exists.
This is consistent with the fact that the arm-to-interarm ratio of the NIR
arm in NGC 5055, 1.3 (\cite{key-Thorn97})  is smaller than that in M 51, 
1.8 -- 3, (\cite{key-Rix93}).  

\begin{figure}
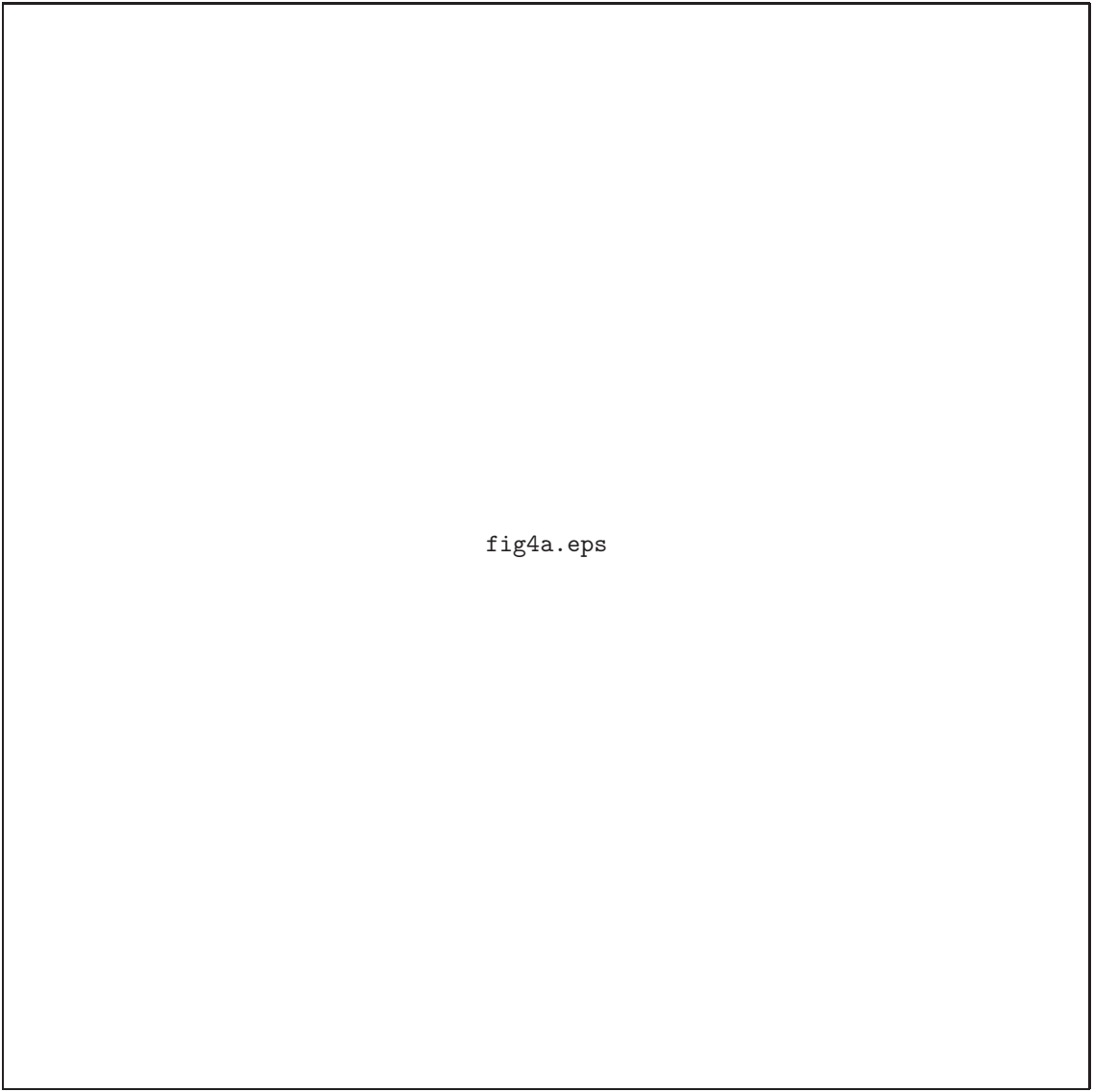

  \begin{center}
    \FigureFile(150mm,150mm){fig4a.eps}
    %%% \FigureFile(width,height){filename}
  \end{center}
  \caption{Velocity channel map. The contour interval and lowest
  contour are 90 mJy km s$^{-1}$beam$^{-1}$, corresponding to the noise
 level. The center velocity of each channel map is indicated at 
 the top-right in each panel.
 The synthesized beam is at the bottom-left in the top-left panel.
 }\label{fig:4a}
\end{figure}

We also find the similar systematic velocity shift in the velocity 
channel maps (figure \ref{fig:4a}a and b).
Each panel of the figure has a velocity range of 2.6 km s$^{-1}$, and  
emissions are seen in each panel elongated along south-west to north-east, namely, 
toward the galactic center of NGC 5055.
In each velocity channel map, the emissions systematically moved 
from west to east with increasing of velocity.
Our results could not show any clear evidence 
for streaming motion caused by a density wave in NGC 5055.

\begin{figure}
  \begin{center}
    \FigureFile(150mm,150mm){fig4b.eps}
    %%% \FigureFile(width,height){filename}
  \end{center}
  \label{fig:4b}
\end{figure}

\begin{figure}
  \begin{center}
    \FigureFile(80mm,80mm){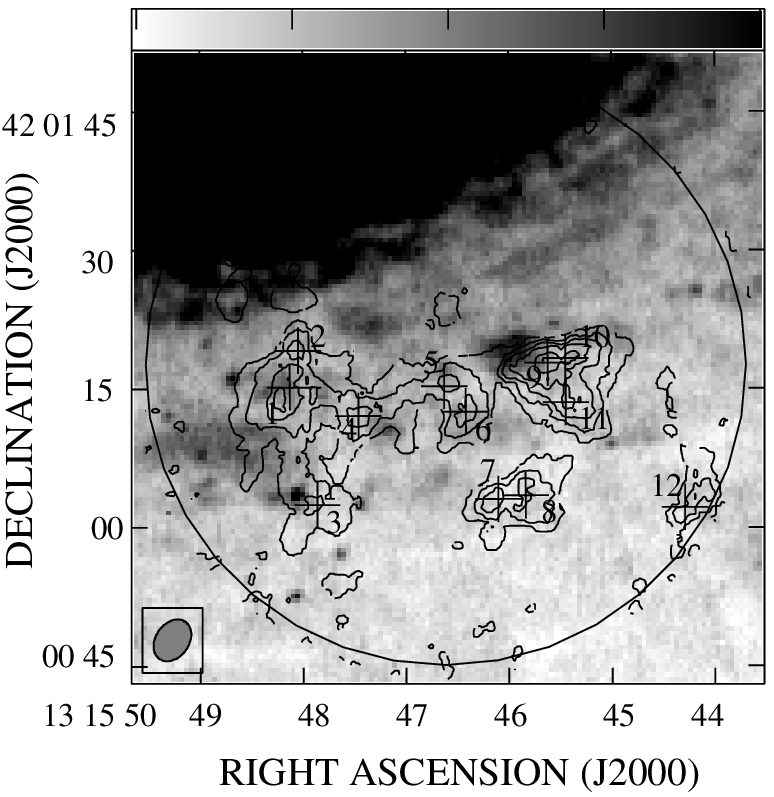} 
\end{center}
  \caption{Individual molecular clouds indicated on the total 
 integrated intensity 
 map and H$\alpha$ image. }
\label{fig:5}
\end{figure}

Figure \ref{fig:5} shows the CO distribution superposed on the H$\alpha$
image obtained by the Subaru telescope.
Although this is not the subtracted continuum (\cite{key-Komiya00}),
no serious problem arises to trace the star forming regions.
In this figure, although some star-forming regions are located near to
the spiral arm, we could not find any systematic offset between the CO and
H$\alpha$ emissions, unlike M 51. 
A clear offset between the CO and H$\alpha$ emissions was found in 
the grand design spiral M 51 (\cite{key-Vog88}), which suggests 
a time delay between gas accumulation due to the density wave and 
the massive star formation traced by H$\alpha$ emission.
In NGC 5055, we could not find such an offset, indicating no systematical 
time delay between the gas accumulation and star formation. 
There are clouds associated both with and 
without massive star-forming regions in both the arm and the interarm
of NGC 5055.

\subsection{Molecular Clouds in the Arm and the Interarm}

\subsubsection{Identification of clouds}

We identified twelve individual molecular clouds
determined by the peaks in the total integrated intensity map.
They are indicated as crosses in figure \ref{fig:5} superposed on the 
total integrated intensity map and H$\alpha$ image. 
Six clouds are distributed on the spiral arm seen in the NIR image
(\cite{key-Thorn97}); five clouds (Nos. 3, 7, 8, 11, and 12)
and one cloud (No. 2) are located on the sides
downstream and upstream of the NIR arm, respectively.

The line profiles of the individual clouds are shown in figure 6.
We must note that the velocity coverage was not complete for five clouds
(Nos. 1, 2, 3, 10, and 11).
Figure 6 also indicates that the identified clouds consist of
a few velocity components, for example, No. 12 cloud.

\begin{figure}
  \begin{center}
    \FigureFile(80mm,80mm){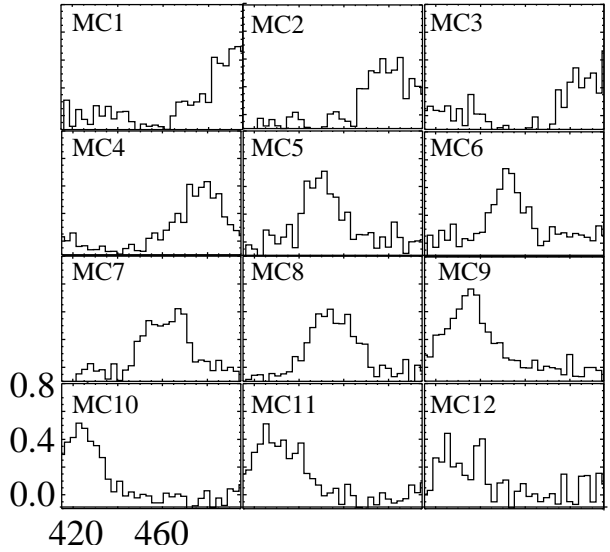} 
\end{center}
  \caption{Line profiles of the individual molecular clouds. The horizontal
 and vertical axes are the velocity (km s$^{-1}$)and intensity (Jy
 beam$^{-1}$), respectively. }
\label{fig:6}
\end{figure}

\subsubsection{Cloud properties}
  
We summarize the properties of individual molecular clouds in 
table \ref{tab:2}.
In the table, 
$D$ and $\Delta v$ are FWHM diameter 
deconvolved with the beamsize and the FWHM velocity width obtained 
by a Gaussian fitting, respectively.
Since Nos. 2 and 9 clouds have a smaller size than the beamsize,
we could not make a deconvolution with the beamsize.
We note that the clouds at the edge of F.O.V. can not cover the
full velocity range due to the limited velocity coverage.
For this reason, the velocity width, CO flux mass,
and virial mass of such clouds were obtained as a lower limit
(Nos. 1, 2, 3, 10, and 11). 
However, most of the clouds seem to cover the velocity range containing
the peak as shown in figure 6, and are considered not to vary by more 
than twice.

We calculated both the CO flux-based mass, $M_{\rm CO}$, and the virial mass, 
$M_{\rm vir}$.
$M_{\rm CO}$ was derived from the CO flux using the CO-to-H$_2$ 
conversion factor in our Galaxy, $X$, $3 \times 10^{20}$cm$^{-2}$ 
(K km s$^{-1}$)$^{-1}$ (\cite{key-Stro88}; \cite{key-Sco87}). 
Assuming a $1/r$ density profile, the virial masses are determined as

\begin{equation}
M_{\rm vir}=99[\Delta V({\rm km~s}^{-1})]^2 D({\rm pc}). 
\end{equation}

In this table, A/I indicates where the cloud was located.
``A'', ``IU'', and ``ID'' mean a cloud in the arm, the upstream side of the arm,
and the downstream side of the arm, respectively. 

The averaged size of individual clouds is 256 pc, which
is larger than the typical size of GMCs in our Galaxy, 
40 pc (\cite{key-Sco87}), but smaller than that of GMAs in M 51(\cite{key-Rand92}),
1 kpc.
The averaged mass, $4 \times 10^6 M_\odot$ is also larger 
than that of the typical GMCs, and corresponds to the mass of 
the largest GMCs in our Galaxy, and
is lower than GMAs in M 51, $\geq 10^7 M_\odot$.

We must point out a possibility that these clouds in NGC 5055 
are aggregations of a small cloud, such as GMCs in our Galaxy.
In this case, the clouds in NGC 5055 can be resolved to some subparts.
However, our information is limited due to the spatial resolution 
of our observations. 
If the clouds in NGC 5055 had smaller subparts, then these would be more 
like GMCs in our Galaxy.

We must also note that these clouds have smaller masses than GMAs seen in 
a previous study by \citet{key-Thorn97}, which have a typical mass of 
$10^7 M_\odot$.
This can be caused by a difference in the spatial resolution 
between our and their observations. 
The map used to identify GMA had a lower spatial resolution,
\timeform{7''}, than ours.
For example, we can see a separation of a few arcsec between
the Nos. 9 and 10 clouds in the total map (figure \ref{fig:5}).
This indicates that each cloud could not be resolved in their map.
It seems reasonable that GMA 10 of \citet{key-Thorn97} 
corresponds to the sum of the Nos. 9 and 10 identified here.  
The sum of the masses for two clouds is $\geq 1.6 \times 10^7 M_\odot$. 
Although this is slightly smaller than the mass of GMA 10 in
\citet{key-Thorn97}, 3.0 or 2.8 $\times 10^7 M_\odot$, we can say that both masses
agree eith each other within the error, considering that this is the lower limit 
due to the limited velocity coverage. 
The fact that the clouds identified here have smaller velocity widths
than those of GMAs by \citet{key-Thorn97} supports that their GMAs consist
of several components such as the clouds identified here.

\begin{figure}
  \begin{center}
    \FigureFile(80mm,80mm){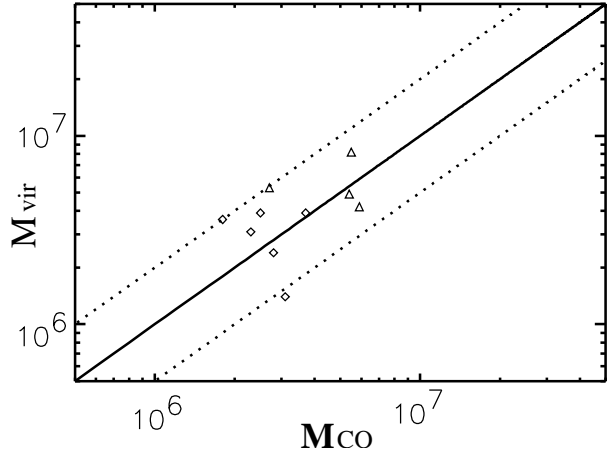} 
\end{center}
  \caption{$M_{\rm CO}$ vs. $M_{\rm vir}$ of the individual molecular
 clouds.
The clouds with a lower limit are indicated as triangles and the rest of
the clouds as diamonds. The solid line shows $M_{\rm CO} = M_{\rm vir}$, and 
 the dotted lines show $M_{\rm CO} = 2 \times M_{\rm vir}$ and 
$M_{\rm CO} = M_{\rm vir}/2$.
}
\label{fig:7}
\end{figure}

We present the $M_{\rm CO}-M_{\rm vir}$ relation of the individual
clouds in figure \ref{fig:7}.  
We find that the $M_{\rm vir}$ of the individual clouds are comparable to
$M_{\rm CO}$ in figure \ref{fig:7} within a factor of $\sim$ 2.
If the $M_{\rm vir}$ is regarded as being a true mass of clouds,
this indicates that we can use the same $X$ value in NGC 5055 
as that of our Galaxy.
The $X$ value depends on the temperature, density, and
metallicity of molecular clouds (e.g., \cite{key-Ari96}; \cite{key-Saka96};
\cite{key-Wil95}).
The $X$ decreases with increasing temperature and increases with density,
and has a larger value in high metallicity than low metallicity
(cf. \cite{key-Wil90}). 
For example, the galactic center in our Galaxy shows a smaller $X$
than that of GMCs in the Galactic plane.
This seems to be reflected to high temperature and high density in
the Galactic center (\cite{key-Dah98}).
Starburst galaxies also have a lower value, indicating that 
they have different properties from our Galaxy (\cite{key-Aal95}),
i.e., a higher temperature and density in the clouds.
On the other hand, high-latitude clouds showed various $X$ (\cite{key-Mag98}).
This may reflect the abundance of CO in the clouds. 
Considering them, the fact that $M_{\rm CO}$ is comparable to $M_{\rm vir}$
indicates that these properties of molecular clouds in
NGC 5055 are not significantly different from those in our Galaxy. 

\begin{figure}
  \begin{center}
    \FigureFile(80mm,80mm){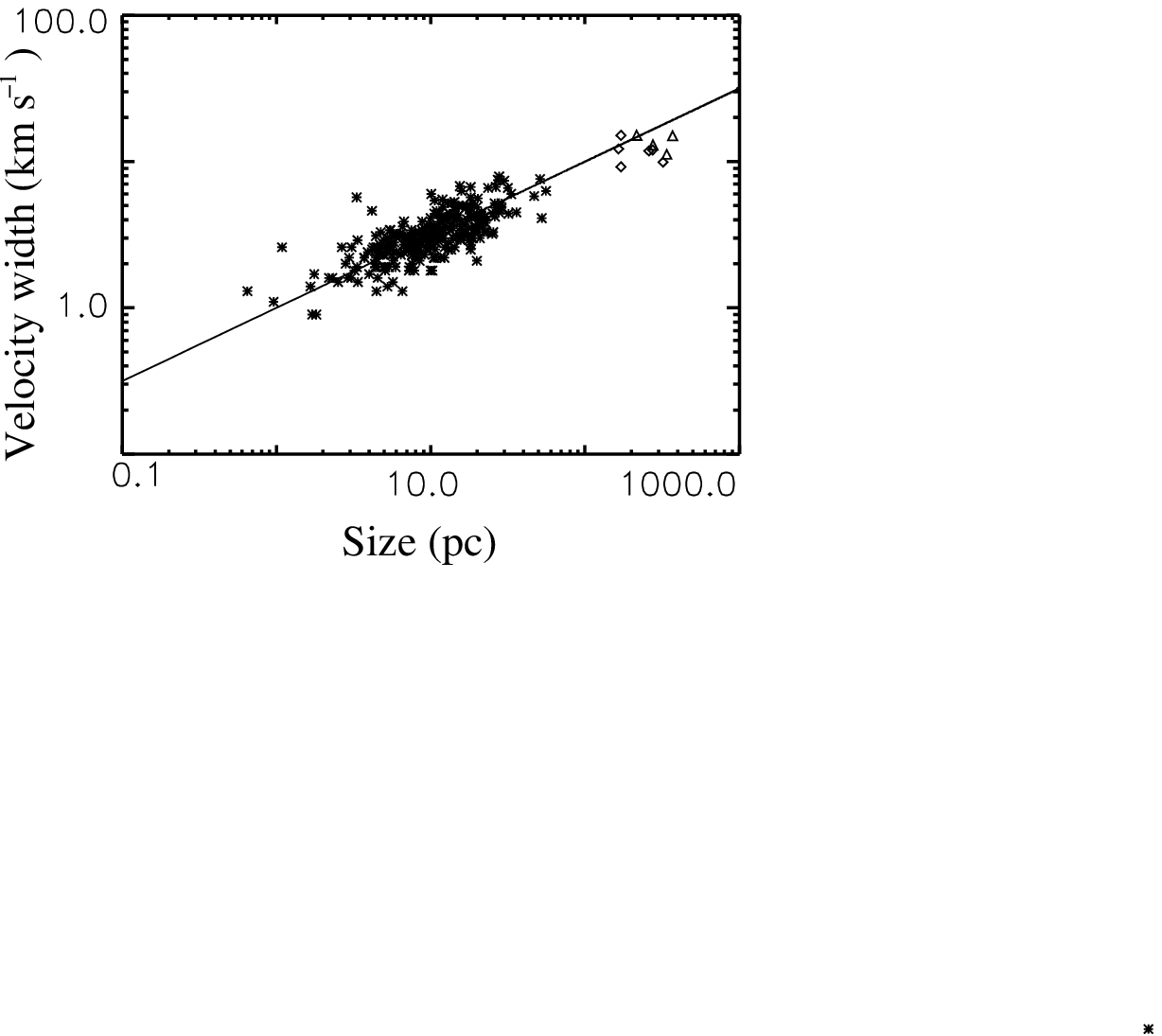} 
\end{center}
  \caption{Size--velocity width relation of the individual GMCs
  in NGC 5055 and Galactic GMCs.
The solid line indicates the relation of $\delta v = D^{0.5}$, which
were obtained for the Galactic GMCs (\cite{key-Sand85}).}
\label{fig:8}
\end{figure}

Figure \ref{fig:8} shows size--velocity relation of individual clouds
along with those of GMCs in our Galaxy (\cite{key-Sand85}). 
This figure indicates that although the individual clouds identified here
have larger masses and sizes than GMCs in our Galaxy,
they are plotted at the same relation as that of GMCs.
This also supports that the molecular clouds in NGC 5055 are similar to
those of GMCs in our Galaxy.
This relation also may support the view that the clouds identified 
here are similar to those of GMCs in our Galaxy. 
Hereafter, we call these clouds GMCs.

Table \ref{tab:2} also gives the averaged values of size, $\Delta v$, $M_{\rm CO}$,
and $M_{\rm vir}$ of the clouds in the arm, and the interarm 
(downstream to arm), respectively.
These values are similar to each other. 
We found
no significant difference between the arm and the interarm.

\begin{table}
  \caption{Properties of the individual molecular clouds.}\label{tab:2}
  \begin{center}
    \begin{tabular}{ccccccccc}
 \hline\hline
No. & $\Delta \alpha$ & $\Delta \delta$ & $D$ & $\Delta v$ & $M_{\rm CO}$ & $M_{\rm vir}$
     & A/I & H\,{\footnotesize II} \\
    & (arcsec) & (arcsec) & (pc) & (km s$^{-1}$) & ($10^6 M_\odot$) & ($10^6 M_\odot$) & & \\ 
\hline \\
1 & -12.5 & -34.1 & 370 & $\geq$15.0 & $\geq$5.5 & $\geq$8.2 & A & H\,{\footnotesize II} \\
2 & -13.4 & -30.2 & *** & $\geq$12.8 & $\geq$1.8 & *** & IU & No \\       
3 & -15.5 & -46.9 & 276 & $\geq$13.9 & $\geq$2.7 & $\geq$5.3 & ID & H\,{\footnotesize II} \\
4 & -19.9 & -37.2 & 259 & 11.8 & 1.8 & 3.6 & A  & H\,{\footnotesize II} \\    
5 & -29.2 & -34.0 & 320 & 9.9  & 2.3 & 3.1 & A  & H\,{\footnotesize II} \\
6 & -31.5 & -36.8 & 171 & 9.2  & 3.1 & 1.4 & A  & ? \\
7 & -35.1 & -46.2 & 165 & 12.2 & 2.8 & 2.4 & ID & No \\ 
8 & -38.0 & -45.8 & 274 & 12.0 & 3.7 & 3.9 & ID & No \\
9 & -40.4 & -31.5 & *** & 12.7 & 10.5& *** & A  & H\,{\footnotesize II} \\   
10& -42.3 & -31.0 & 338 & $\geq$11.2 & $\geq$5.9 & $\geq$4.2 & A  & ?\\ 
11& -42.4 & -35.8 & 216 & $\geq$15.1 & $\geq$5.4 & $\geq$4.9 & ID & No\\
12& -55.2 & -47.1 & 171 & 15.1 & 2.5 & 3.9 & ID & No \\    
\\
\\
average & & & 256 & 12.8 & 4.2 & 4.1 & & \\
A       & & & 292 & 11.6 & 4.9 & 4.1 & & \\ 
ID      & & & 220 & 13.7 & 3.4 & 4.1 & & \\
\hline

\end{tabular}
\end{center}
\end{table}

\section{Discussion}

\subsection{GMC formation in the NGC 5055}

As mentioned above, there is no difference in the properties of the GMCs  
between the arm and the interarm in NGC 5055.
We also see no systematic offset between the spiral arms of the molecular gas 
and H$\alpha$ in NGC 5055.

Here, we consider where the GMCs seen in the interarm  were
formed, namely in the arms or in the interarm.
For this purpose, we compare two timescales,
i.e., the lifetime of the clouds and the traveling time of the clouds from 
the arm to the interarm, as follows.

First, we estimated the traveling timescale of the GMCs, i.e., timescale 
after leaving arm.
For example, the GMC seen in the interarm of NGC 5055, 
No.\ 7, has a distance of 3.3
kpc from the arm based on a face-on view.
Assuming the rotational velocity as 200 km s$^{-1}$, the timescale was
estimated to $1.6 \times 10^7$ years.
This is the lower limit because the spiral arm pattern also 
moves in the same direction.

Next, we estimated the lifetime of the cloud.
Since the time lag of the CO and H\,{\footnotesize II} region seen 
in M 51 is $10^7$ years
and there are no GMCs associated with the H\,{\footnotesize II} regions, 
it is thought that the lifetime of the cloud is shorter than this timescale.
The timelag of the CO and H\,{\footnotesize II} regions in NGC 6951 
was also evaluated to be $10^6$ years (\cite{key-Kohno99}). 
It is therefore considered that the lifetime of the clouds are shorter than 
the traveling time from the arm, and that the interarm GMC was
formed in the interarm.

However, there is another way to estimate the lifetime of the cloud.
\citet{key-Elme00} suggested that 
the lifetime of cloud  was estimated to be 
comparable with two times dynamical time of a cloud
from observations.
The dynamical timescale is $\tau_{\rm dyn} = (\frac{4}{3}\pi R_{\rm cl}^3/GM_{\rm cl})^{1/2}$,
where $R_{\rm cl}$ and $M_{\rm cl}$ are the radius and the mass of the cloud, 
respectively.
Adopting values of No.¥ 7 GMC as $R_{\rm cl}$ and $M_{\rm cl}$, 
we obtained the lifetime as $2\times\tau_{\rm dyn} \sim  2.6\times10^7$ years.
It seems reasonable that the lifetime of the cloud is similar to the 
above traveling time.
Therefore, it is difficult to see whether the interarm GMCs
were formed in the arm or in the interarm. 
In order to clarify this situation, it is necessary to increase the number 
of samples and to investigate the relationship between the molecular gas and 
the star-forming regions.

\subsection{Flocculent vs. Grand Design}

Here, we recall no clear systematic offset with the H\,{\footnotesize II} 
region, which is seen in the arm of M 51.
This offset indicates that the star formation started on the arm 
at the same time by a density wave in M 51.
In other words, the absence of an offset suggests that 
the star formation may not have
necessarily started at the same time in NGC 5055.
We could also find no GMAs on the arm in NGC 5055, unlike the case of M 51.
From these facts, we suggest that the density wave in NGC 5055  plays 
no direct role in either GMA formation or in the star formation;
though a density wave plays a role in the accumulation of molecular gas in arms, 
arm-to-interarm ratio of molecular gas is not high compared with that of the 
grand design spiral M 51.

The gravitational instability of a gas disk plays 
an important role in star and cloud formation
(\cite{key-Ken89}; \cite{key-Elme94};
\cite{key-Tosa97}; \cite{key-Shio98}; \cite{key-Kohno02}). 
It is reasonable that they easily occur on the arm, 
because the density in the arm is higher than that in the interarm.
From these facts, it is possible to build up a hypothesis as follows.
In the flocculent galaxy NGC 5055, since the arm-to-interarm ratio of CO is low,
$\sim 2$, the difference in the gas density between the arm and the interarm 
is not significantly large. 
Therefore, the formations of clouds and stars could occur not only in the arm
but in the interarm due to a local fluctuation of the gas density, though
it may occur more easily in the arm. 

On the other hand, the grand design spiral M 51 shows a clear offset between
the molecular gas and the star-forming regions, though we can find the some
star-forming regions in the interarm.
M 51 also shows a difference of the 
molecular clouds between the arm and the interarm (see subsection 3.1).
These facts suggest that a strong density wave has an effect on the
GMA and the star formation in the case of M 51, unlike NGC 5055.
Since the gas density becomes high in the arm due to the accumulation of gas 
by a strong density wave,
we consider the following three scenarios.
At first, the gas density is higher than the critical density for a 
large-scale instability, and GMA formation occurs due to the instability
in the arms of M51 (\cite{key-Elme94}). 
Once GMAs are formed,
star formation may follow in the GMAs, as suggested by a comparison
of the scale between large-scale instability and star-forming
complexes (\cite{key-ElmeD94}).
Second, GMAs may be formed by the collision of molecular clouds in
the arm (\cite{key-Kwan87}). 
Cloud collisions must also trigger star formation (\cite{key-Lars88})..
Third, both mechanisms mentioned above work simultaneously.

In a flocculent galaxy, although a weak density wave accumulates gas,
the density of the gas becomes not high 
and not enough for large scale instability or/and collision to occur 
and trigger star formation.
This may produce a different morphology of flocculent and grand design
spiral arms.   
To sum up, in grand design spiral galaxy with a strong density wave, 
since the star formation is triggered by a strong density wave in the arm, 
the optical arm with offset from the dust lanes clear,
while the star formation in the arm of flocculent galaxy with a weak density
wave is not very much. 
Therefore, we can see no clear optical spiral arm in the flocculent galaxy. 
We thus propose that the strength of density wave may control the arm 
morphology.

This is the first example, 
and we don't yet know whether our findings for NGC 5055 are also applicable to 
other flocculent galaxies or not.
To confirm this scenario, the observation in a large area of NGC 5055
and observations in other galaxies will be needed for a verification.
In particular, it will be necessary to observe galaxies with various
arm classes and to investigate the relation between the amplitude of the density
wave, and the arm morphology.

\section{Conclusions}

We present the results of high-resolution ($\sim \timeform{4''}$) 
$^{12}$CO($J$~=~1--0) mapping observations toward
the southern bright arm region of the nearby spiral galaxy NGC 5055 
carried out with NMA. 
The velocity resolution of 2.6 km s$^{-1}$ was very high in 
extragalactic observations.
The main conclusions are summarized as follows:

1. The molecular clouds are mainly distributed along 
a spiral arm seen in near-infrared observations. 
The arm consists of several clumpy structures whose typical size and mass
are a few 100 pc and 10$^6 M_\odot$, respectively. 
The values are comparable to those of GMCs in our Galaxy, and
smaller than GMAs in M 51.
There is no offset between the molecular spiral arm and the H$\alpha$
emission seen in the grand design spiral M 51.

2. We also detected interarm molecular clouds as well as on arm.
Their masses and sizes are similar to those on the arm.
This differs from grand design spiral M 51, which has GMAs on the arm
and no GMAs in the interarm.
It is difficult to see whether these interarm clouds were formed in the arm 
or the interarm based on the argument of the lifetime of clouds.   
  
3. The virial masses of the clouds agree well with the CO flux masses
within a factor of 2.
We can, therefore, use the CO-to-H$_2$ conversion factor in our
Galaxy, indicating that the properties of clouds, such as the temperature
and the density, are not significantly different.
The size--velocity width relation was also plotted as the same relation as 
the Galactic GMCs, supporting this idea.

4. No existence of GMAs and no clear systematic offset between the molecular gas 
and H\,{\footnotesize II} regions suggest the view that, in a flocculent galaxy, 
cloud formation, and the following star formation occur in both the arm and 
the interarm due to an enhancement of the gas by a local fluctuation.

\vspace{0.5cm}

Acknowledgment

We are grateful to the NRO staff for operating and 
improving of the NMA.
We also thank to Drs. Y. Komiyama and H. Fukushima for offering the 
optical data and giving helpful comments about them.
We would like to thank Dr.\ K. Kohno for useful comment and discussion.
Y.S. and K.N. thanks the Japan Society for Promotion of Science (JSPS)
Research Fellowships for Young Scientists. 
%%%%%%%%%%%%%%%%%%%%%%%%%%%%%%%%%%%%%%%

%%%
% See the manual for the detail.
%%%

\end{document}